\documentclass{egpubl}
\usepackage{eurovis2025}

\SpecialIssuePaper         % uncomment for final version of Computer Graphics Forum, special issue
\CGFccby
\usepackage[T1]{fontenc}
\usepackage{dfadobe}  

\usepackage{cite}  % comment out for biblatex with backend=biber
\BibtexOrBiblatex
\electronicVersion
\PrintedOrElectronic
\ifpdf \usepackage[pdftex]{graphicx} \pdfcompresslevel=9
\else \usepackage[dvips]{graphicx} \fi

\usepackage{egweblnk}
\usepackage{xcolor}
 
\title[Designing Reality-Based VR Interfaces for Geological Uncertainty]%
      {Designing Reality-Based VR Interfaces for Geological Uncertainty}

\author[R. Mota et al.]
{\parbox{\textwidth}{\centering R. Mota$^{1}$, E. Sharlin$^{1}$,
        and U. Alim$^{1}$ 
        }
        \\
{\parbox{\textwidth}{\centering $^1$Department of Computer Science, University of Calgary, Canada\\
       }
}
}
\usepackage{amsfonts}
\begin{document}

% uncomment for using teaser
% \teaser{
%  \includegraphics[width=0.9\linewidth]{eg_new}
%  \centering
%   \caption{New EG Logo}
% \label{fig:teaser}
%}

\maketitle
%-------------------------------------------------------------------------
\begin{abstract}
   Inherent uncertainty in geological data acquisition leads to the generation of large ensembles of equiprobable 3D reservoir models. 
   Running computationally costly numerical flow simulations across such a vast solution space is infeasible. A more suitable approach is to carefully select a small number of geological models that reasonably capture the overall variability of the ensemble.
   Identifying these representative models is a critical task that enables the oil and gas industry to generate cost-effective production forecasts.
   Our work leverages virtual reality (VR) to provide engineers with a system for conducting geological uncertainty analysis, enabling them to perform inherently spatial tasks using an associative 3D interaction space.
   We present our VR system through the lens of the reality-based interaction paradigm, designing 3D interfaces that enable familiar physical interactions inspired by real-world analogies---such as gesture-based operations and view-dependent lenses.
    We also report an evaluation conducted with 12 reservoir engineers from an industry partner. Our findings offer insights into the benefits, pitfalls, and opportunities for refining our system design. We catalog our results into a set of design recommendations intended to guide researchers and developers of immersive interfaces---in reservoir engineering and broader application domains.

%-------------------------------------------------------------------------
%  ACM CCS 1998
%  (see https://www.acm.org/publications/computing-classification-system/1998)
% \begin{classification} % according to https://www.acm.org/publications/computing-classification-system/1998
% \CCScat{Computer Graphics}{I.3.3}{Picture/Image Generation}{Line and curve generation}
% \end{classification}
%-------------------------------------------------------------------------
%  ACM CCS 2012
%(see https://www.acm.org/publications/class-2012)
%The tool at \url{http://dl.acm.org/ccs.cfm} can be used to generate
% CCS codes.
%Example:
\begin{CCSXML}
<ccs2012>
<concept>
    <concept_id>10010147.10010371.10010352.10010381</concept_id>
    <concept_desc>Visualization application domains~Visual analytics</concept_desc>
    <concept_significance>300</concept_significance>
</concept>
<concept>
    <concept_id>10010583.10010584.10010587</concept_id>
    <concept_desc>Physical sciences and engineering~Earth and atmospheric sciences</concept_desc>
    <concept_significance>100</concept_significance>
</concept>
</ccs2012>
\end{CCSXML}

\ccsdesc[300]{Visualization application domains~Visual analytics}
\ccsdesc[100]{Physical sciences and engineering~Earth and atmospheric sciences}

\printccsdesc   
\end{abstract}

%-------------------------------------------
% Introduction
%-------------------------------------------
\section{Introduction}
%\vspace{-5px}

A petroleum reservoir is a piece of earth located deep underground. Its digital version is a 3D reservoir model, whose structure often employs corner point geometry since this can represent complex reservoir features (\textit{e.g.}, faults, fractures, horizons), \textcolor{black}{as seen in Fig.~\ref{fig:reservoirModel}}. A model typically consists of thousands to millions of cells, with various associated properties, such as porosity, permeability, and oil saturation. As reservoirs lie hundreds to thousands of feet in the subsurface, data acquisition is challenging and restricted to data sampled from sparse locations of the reservoir field. 
Reservoir data is thus wrapped in uncertainty about the spatial distribution of reservoir properties. To model property uncertainty, geostatistical stochastic simulations are used to generate several (possibly hundreds or thousands) alternative, equally-probable models of the spatial distribution of that property \textcolor{black}{\cite{ma2011}}. These are called \textit{geological realizations}. 
Due to resource limtations, performing fluid flow simulations in all geological realizations is unfeasible. A more practical approach is to carefully select a small set of models that reasonably represent the overall uncertainty, and then run flow simulation only on this small-best subset. We call these selected realizations \textit{representative models} of the original reservoir ensemble.

Industry's standards to select representative models are automated techniques like ranking, random selection, and probability-based methods \textcolor{black}{\cite{Ballin:1992, Scheidt:2010}}. These automated methods are costly due to the computation of a large number of grid cells, and employ global aggregation strategies that fail to capture existing local heterogeneities in the realization models. 
Recently, Sahaf \textit{et al.} \textcolor{black}{\cite{sahaf2016}} proposed the first interactive selection process: engineers first define a region of interest on the reservoir model; a clustering algorithm then groups the realization ensemble, accounting only for this particular area; afterwards, a few models from each cluster are selected, thereby yielding to the representative models. 
This user-guided framework takes advantage of engineers' knowledge to enhance representative set selection: it captures local variability relevant to experts, and decreases computational cost by considering only a portion instead of the entire grid geometry. This analytical framework is used as the underlying basis for our system.

\vspace*{-5px}
\noindent Furthermore, Sahaf \textit{et al.}~\cite{Sahaf:2018} argue that the assessment of geological uncertainty is inherently exploratory and heavily reliant on human input and judgment throughout the process. However, although the strength of their framework lies in the interplay between engineers and 3D reservoir models, it still lacks adequate user support for performing fundamental spatial tasks---\textit{e.g.}, inspecting both external and internal cells to determine whether to marke them as relevant.
Our work builds upon this analytical framework by leveraging virtual reality (VR) to develop intuitive and expressive 3D interactions intended to reduce user effort in tasks associated with geological uncertainty analysis. We operate on the premise that grounding interactions in users’ pre-existing real-world knowledge and motor skills can reduce cognitive load, as users already possess the capabilities required for system operation. 

\noindent \textbf{Contribution} Hence, the real world served as the foundation for designing and contributing a \textbf{{VR system}} tailored for uncertainty analysis of geological realizations. Our design strives for fluid interfaces \textcolor{black}{\cite{Csikszentmihalyi:2008, Elmqvist:2011}}, leveraging VR technology to enable familiar physical transactions with 3D interfaces for reservoir analysis.
This paper also contributes a user evaluation conducted with 12 reservoir engineers from an industry partner. The findings support our claims, provide ideas for refinements, and suggest VR as a promising alternative for engineers.

\noindent \textbf{Outline} This paper is structured as follows: Section \ref{sec:relatedWork} reviews related work on conventional and spatial interfaces for reservoir analysis. Section \ref{sec:systemDataModel} outlines the system's data pipeline, and Section \ref{sec:systemDesign} details its design. Section \ref{sec:userStudy} describes a user evaluation conducted with 12 reservoir engineers, while Section \ref{sec:resultsDiscussion} reports and discusses the findings, deriving design opportunities for future research on immersive interfaces. Finally, Section \ref{sec:conclusionFutureWork} concludes the paper.

%-------------------------------------------
% Related Work
%-------------------------------------------
%\vspace{-6px}
\section{\textcolor{black}{Related Work}}
\label{sec:relatedWork}
%\vspace{-3px}

This section reviews existing commercial solutions and research efforts for reservoir exploration and analysis, considering both conventional and immersive mediums.

\begin{figure}[t!]
\centering
\includegraphics[width=1\linewidth]{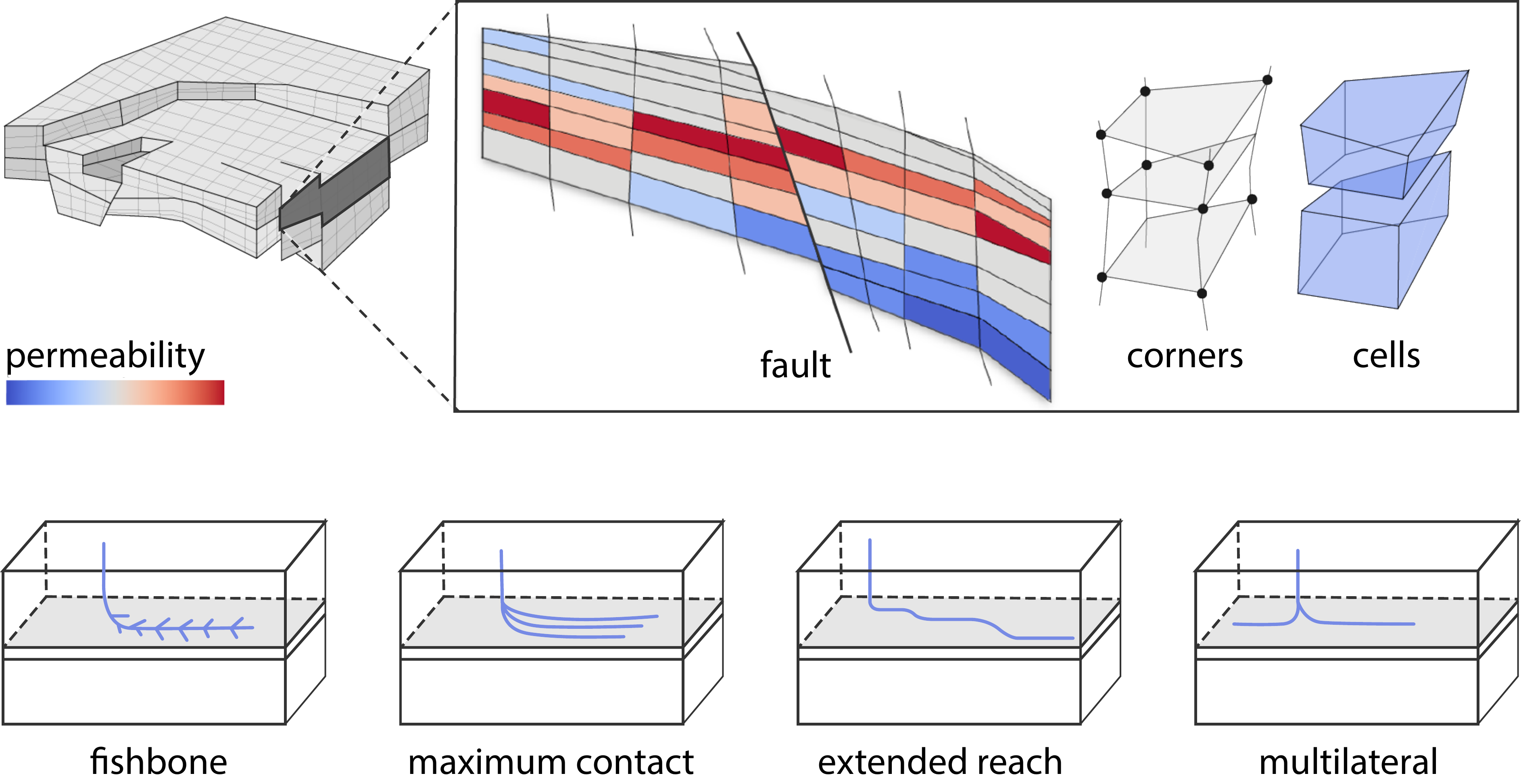}
\vspace*{-4px}
\caption{Illustration of a reservoir model (\textit{top}) and common well profiles (\textit{bottom}). Corner-point gridding is used by the oil industry as it can represent reservoir features (e.g., faults, fractures) by specifying eight corners of each grid cell. Each cell can be associated with static and dynamic attributes (e.g., rock type, permeability, and oil saturation).}
\label{fig:reservoirModel}
\vspace*{-0.4cm}
\end{figure}

\begin{figure}[b!]
\centering
%\vspace{-0.3cm}
\includegraphics[width=1\linewidth]{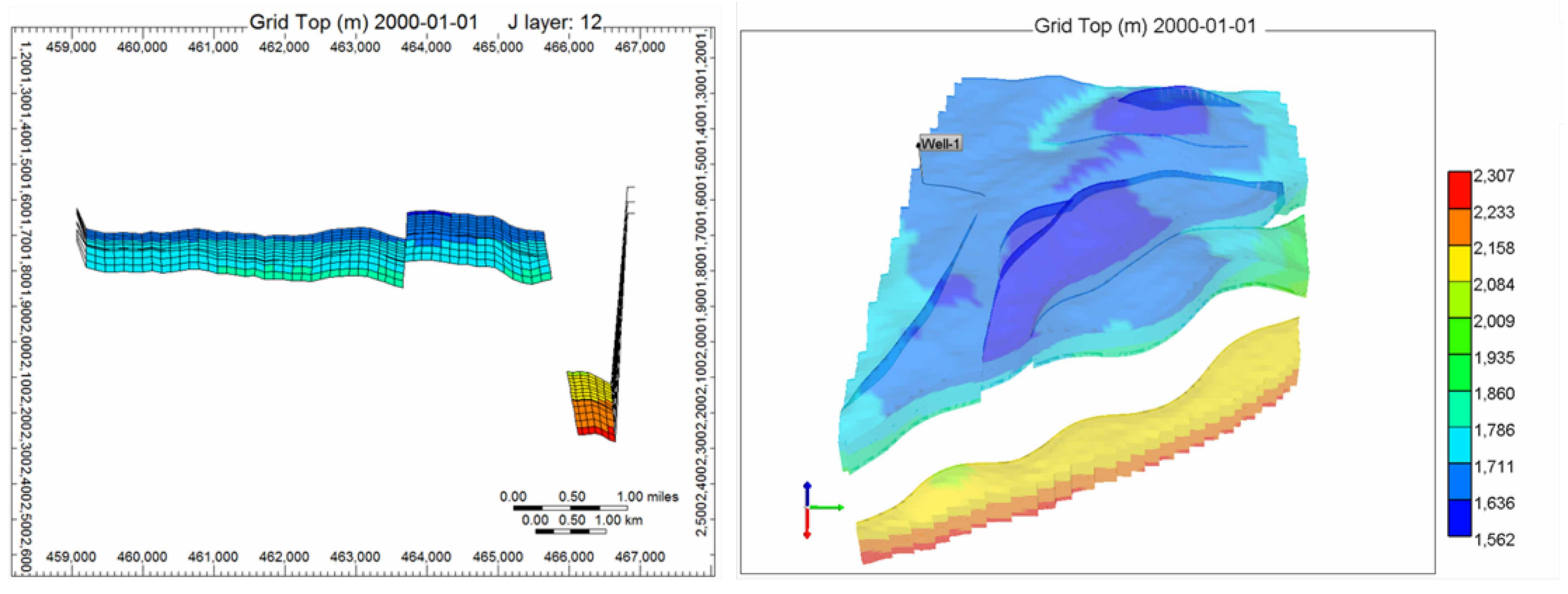}
%\vspace*{-4px}
\caption{Cross-section and $45^\circ$ view on CMG Suite\textcolor{black}{\cite{CMG}}.}
\label{fig:cmg}
\vspace{-0.2cm}
\end{figure}

\noindent \textbf{Commercial software} Although the oil sector relies heavily on the analysis of 3D reservoirs, current commercial solutions offer limited visualization and interaction capabilities \textcolor{black}{~\cite{Petrel, CMG}}. To illustrate, to create a well trajectory, engineers typically set an axis-aligned cross-section~(\textcolor{black}{Fig. \ref{fig:cmg}--\textit{right}}), select the first perforation cell on the topmost layer of the reservoir, iteratively adjust the sectional plane to select subsequent perforation cells, and finally visualize the resulting well in a 3D view \textcolor{black}{~(Fig. \ref{fig:cmg}--\textit{left})}. 
This workflow presents at least two limitations: first, it restricts interaction to axis-aligned sections; second, it lacks contextual awareness, requiring users to continuously switch between sectional views to correlate data---especially with non-conventional wells~(\textcolor{black}{Fig. \ref{fig:reservoirModel}--\textit{bottom}}).

\noindent \textbf{Conventional displays} Due to the limited support from existing systems, there have been efforts in reservoir engineering and geosciences towards novel techniques to improve the ease, speed, and accuracy of reservoir analysis.
Most approaches target 2D interface technologies: classic desktop workstations \textcolor{black}{\cite{martins2015, 2016:ICO}} and touch-enabled devices like tabletops \textcolor{black}{\cite{Somanath:2014, Sultanum:2011}}. Amorim \textit{et al.} introduced a desktop-based sketching system for modeling 3D geological structures, allowing users to draw multistroke symbols on a 2D canvas that are translated into 3D arrangements of rock layers (\textit{e.g.}, dips, domes, basins etc.) \textcolor{black}{\cite{Amorim:2014}}. Somanath \textit{et al.} explored techniques to create well paths on tabletops; in one approach, users first select and orient an axis-aligned plane using a graphical widget and then draw the well trajectory along that plane using their fingers \textcolor{black}{\cite{Somanath:Thesis:2012}}. Both touch- and mouse-based interaction face a fundamental problem: the inherent challenge manipulating 3D data using 2D interaction spaces. This may lead to non-intuitive mappings that place a high cognitive load on users to perform 3D tasks.
For instance, the sketch-based system proposed by Amorim \textit{et al.} relies heavily on users' ability to recall how 2D symbols map to 3D modeling operations, and to mentally foresee the resulting 3D geological structures \textcolor{black}{~\cite{Amorim:2014}}. This leads to a system more difficult to learn and to use.
Similarly, as part of their analytical framework, Sahaf \textit{et al.} proposed a desktop system resembling commercial software and suffering from similar limitations---\textit{e.g.}, in a 2D view, users must use a lasso tool to select a surface reservoir area, which is then extruded along the view plane to form a 3D volume of interest  \textcolor{black}{~\cite{Sahaf:2018}}.

\noindent \textbf{Immersive displays} In response to the above limitations, there has been increasing interest in integrating 3D interfaces into geoscience and engineering workflows \textcolor{black}{\cite{Lidal:2007, kinsland2015, Mota:2016}}. 
Ragan \textit{et al.} investigated small-scale spatial judgment tasks, which requires careful inspection of objects that are small relative to the scale of their surrounding environment \textcolor{black}{\cite{ragan2012studying}}. They evaluated the effects of field-of-regard, stereoscopy, and head-tracked rendering on performance (\textit{e.g.}, time and number of errors) for identifying collisions and gaps in complex underground cave systems. Their findings suggest that higher-fidelity immersive can improve user performance. 
Gruchalla proposed a system for editing well paths in mature oil fields, which include tortuous, intertwined underground wells \textcolor{black}{\cite{Gruchalla:2004}}. By comparing task completion times and correctness between a CAVE-like environment and a desktop, authors found speed and accuracy improvements in the immersive space.

\begin{figure*}[t!]
\centering
\includegraphics[width=1\linewidth]{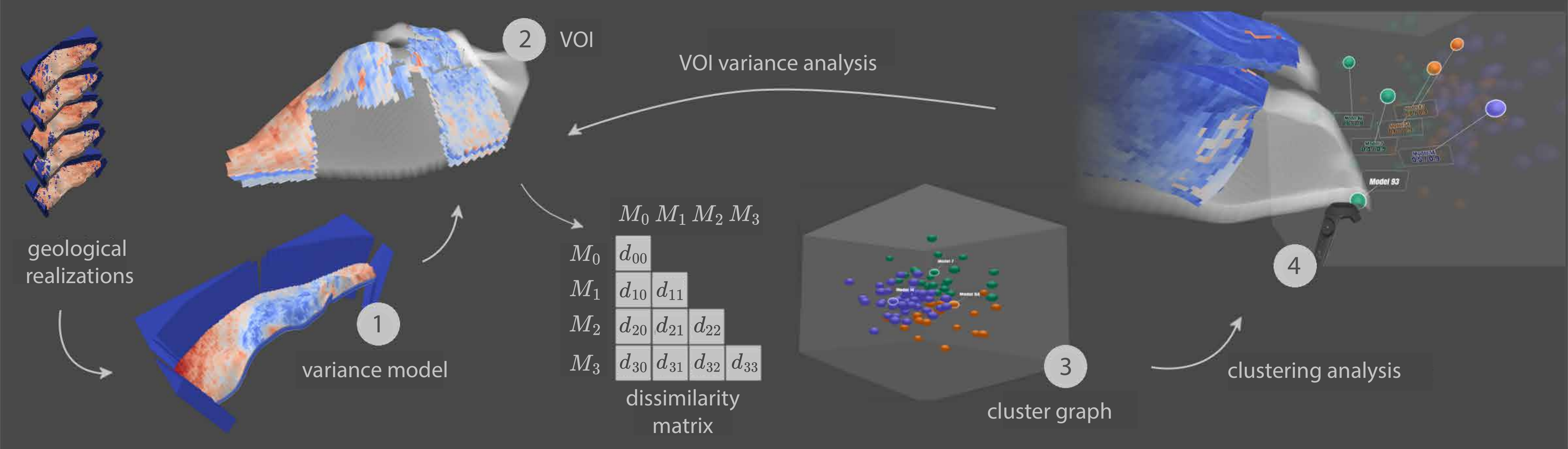}
%\vspace{-0.6cm}
\caption{Geological uncertainty analysis workflow. (1) Color-coded variance model from the input reservoir ensemble. (2) VOI selection. 
(3) VOI-based clustering produces color-coded clusters and a default representative set. (4) Clustering analysis identifies additional samples.}
\label{fig:workflow}
%\vspace{-0.55cm}
\end{figure*}

\noindent The aforementioned studies showcase how spatial interfaces can offer a more effective medium for viewing and interacting with intricate data common in the oil industry, such as geological models, drilling wells, and fluid flow. Our work addresses a specific need in this field: the uncertainty analysis of geological realizations. To our knowledge, it is the first to incorporate themes of realism and physicality to enable familiar interactions with interfaces for reservoir visualization and analysis.

%-------------------------------------------
% System Data Model
%-------------------------------------------

%\vspace{-6px}
\section{\textcolor{black}{System Data Model}}
\label{sec:systemDataModel}
%\vspace{-3px}

This section details the foundation for our VR system: the framework for geological uncertainty analysis recently proposed by Sahaf \textit{et al.} \cite{sahaf2016}. Regular meetings with engineers, working collaboratively with us, were essential for us to acknowledge the steps involved in uncertainty analysis. The workflow consists of a series of steps required to accomplish the goal of \textit{selecting representative models} from a reservoir ensemble---see \textcolor{blue}{Fig. \ref{fig:workflow}}. Each step is described below.

\noindent \textbf{1.} The user inputs i) a set of geological realizations, and ii) one or more properties used to compute per-cell variance. This metric quantifies uncertainty: higher variability indicates greater property uncertainty. Importantly, if the input properties are strongly correlated with fluid flow, then areas of high variance correspond to regions where flow simulation results are likely to differ significantly across realizations. The output of this step is a single model whose cells are associated with variance values---referred to as the \textit{variance model}.

\vspace{-2px}
\noindent \textbf{2.} The user selects a volume of interest (VOI) in the variance model.

\vspace{-2px}
\noindent \textbf{3.} A similarity value is computed for each pair of geo realizations, based on the mutual information \textcolor{black}{\cite{Goshtasby:2012}} shared between their VOI cells. These pairwise similarity values are used as input to a multi-dimensional scaling (MDS) method \textcolor{black}{\cite{France:2011}}, which projects the realizations into points in a 3D space. MDS reduces the original high-dimensional space to a lower, projected dimension, where the absolute 3D positions are not meaningful---the relative distances between points, however, are. These distances quantifies similarity between geo realizations: the closer the points, the more similar the models.
Next, the distance values are used in a kernel $k$-means clustering algorithm \textcolor{black}{\cite{Dhillon:2004}} to group similar realizations. The result is a \textit{3D graph}, where each node represents a geo realization and the pairwise distances between nodes encode their similarity. Nodes are allocated into clusters, and each cluster has a \textit{center node}---the realization that best represents the variability within the cluster.

\vspace{-2px}
\noindent \textbf{4.} By default, the center nodes of each cluster are selected to compose the initial representative set. At this point, variance within the VOI is computed based only on the selected nodes---\textit{i.e.}, the current representatives. Users may then iteratively refine this set by selecting additional sample nodes through a two-step process: \textit{i}) \textit{node distance analysis} and \textit{ii}) \textit{VOI variance evaluation}. 
First, users inspect the graph for outliers---\textit{i.e.}, nodes located far from the cluster centers, and potentially isolated from neighboring realizations. Second, users may select such outliers, triggering a recomputation of the VOI variance to include the newly added node. They can then assess alterations in the variance distribution to decide whether the updated better reflects reservoir heterogeneity. If so, the node is retained as part of the representative set. This iterative process continues as resources permit, with the final set selection used for subsequent flow simulation studies.

\noindent Sahaf \textit{et al.} \cite{Sahaf:2018} argue that this workflow enhances uncertainty analysis by incorporating domain expertise. Engineers can designate critical local regions of the reservoir (\textit{e.g.}, a high-variance zone near a wellbore or aquifier) as a basis for clustering realizations.
This \textit{local} (VOI-based) approach is preferred over a \textit{global} clustering, which aggregates data across the entire reservoir and may obscure relevant local variations. Another benefit is reduced processing time: by discarding a number of grid cells that do not significantly contribute to reservoir variability, the system offers more responsive, real-time analysis. Our system builds upon the aforementioned analytical framework, proposing 3D interfaces to facilitate this local, expert-guided exploration and analysis.

%-------------------------------------------
% System Design
%-------------------------------------------
%\vspace{-6px}
\section{System Design \& Implementation}
\label{sec:systemDesign}
%\vspace{-3px}

This section builds upon the aforementioned data pipeline and describes our VR system design, outlining its design goals and the rationale behind spatial interfaces inspired by real-world analogies.

%------------------------------------------------
% Design Goals
\vspace*{-7px}
\subsection{Design Goals}
%\vspace{-3px}

To forge fluid interfaces, we established the following design goals:

\noindent \textbf{DG\textsubscript{\textit{Real.}}} \textit{Evoke realism as necessary}.
We intend to exploit the reality-based interaction paradigm\textcolor{black}{~\cite{Jacob:2008, Hutchins:1985}}, employing themes from the real world---\textit{e.g.}, na\"{i}ve physics, bodily awareness, and environmental cues---to allow users to leverage their existing real-world knowledge. This design strategy aims at reducing user effort to memorize gesture vocabularies before becoming proficient with the system. 
However, while realism serves as a key design inspiration, we intentionally incorporate non-realistic elements when necessary to support other design goals, including: ergonomics, to reduce the risk of repetitive stress injuries and fatigue; efficiency, to improve task speed and accuracy; and expressive power, to support a broader range of operations. 

\vspace{-2px}
\noindent \textbf{DG\textsubscript{\textit{VR.}}} \textit{Optimize for VR.} We focus on designing for the emerging generation of virtual reality, which we consider particularly well-suited for our application domain due to its inherently volumetric nature, intricate geometry, and interconnecting layers. 
Our design leverages core VR capabilities: head- and hand-tracking to perform compound spatial tasks; stereoscopy and motion parallax to enhance depth perception and spatial understanding; and kinesthesia for body-relative interactions---\textit{e.g.}, direct manipulation, physical mnemonics, and gestural input. At the same time, we intentionally avoid interactions known to be problematic in 3D spaces---\textit{e.g.}, handle small objects or rely on high-precision input.

\vspace{-2px}
\noindent \textbf{DG\textsubscript{\textit{Scale.}}} \textit{Support scalability}.
Our goal is to visualize reservoirs with high complexity and arbitrary grid topologies at interactive frame rates on standard graphics workstations. In this context, the combined demands of high resolution and high frame rate requirements in VR pose a greater challenge than traditional desktop visualization, where lower frame rates and intermittent pauses for computation or data loading are generally more acceptable. To address this, we leverage GPU capabilities to support real-time physical interactions, encouraging users to shift from cognitively demanding operations to perceptual-level engagement---such as relying on muscle memory\textcolor{black}{~\cite{Bartram:1997, Grossman:2001}}.

\noindent The aforementioned guidelines incorporate design principles intended to help users maintain a state of flow during their work, transforming the sense-making process into a more efficient and even enjoyable experience~\cite{Csikszentmihalyi:2008, Elmqvist:2011}. Productivity can be improved through: \textit{i}) reducing task steps, as users’ familiarity with their own body and environment enables the combination of multiple actions into a single gesture; \textit{ii}) lowering cognitive load and learning time by employing interaction metaphors grounded in real-world experiences; and \textit{iii}) minimizing context switching, allowing users to keep their attention focused on the task at hand.

%------------------------------------------------
% Spatial Transformations via Physical Metaphors

\begin{figure}[t!]
\centering
\includegraphics[width=0.85\linewidth]{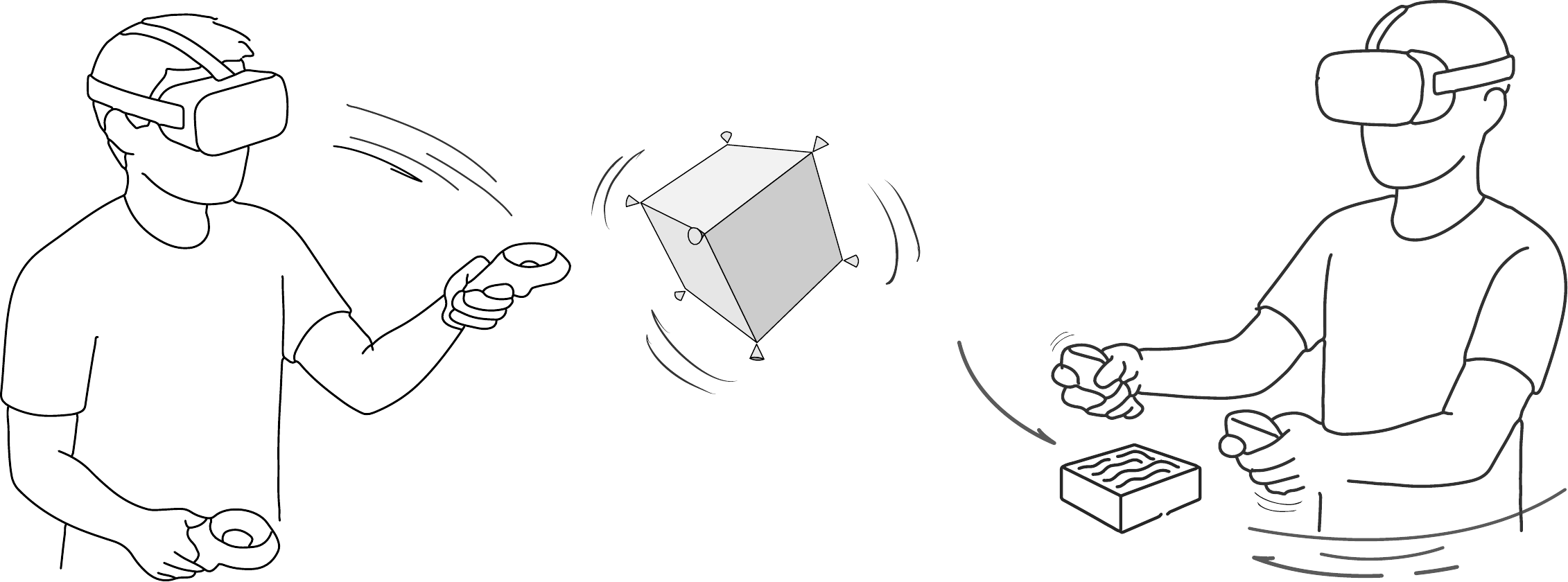}
%\vspace{-0.2cm}
\caption{Deletion (left) and scaling (right) operations are grounded in universally recognizable gestures, designed to promote rapid learnability with low cognitive load, potentially enabling eyes-free execution.
}
\label{fig:gestures}
\vspace{-3px}
\end{figure}
\vspace*{-5px}

%\[
%\mathcal{P} = \left\{ \vec{x} \in \mathbb{R}^3 \ \middle| \ (\vec{x} - \vec{c}) \cdot \vec{a} = 0 \right\}
%\]
\begin{figure}[b!]
\centering
\includegraphics[width=0.75\linewidth]{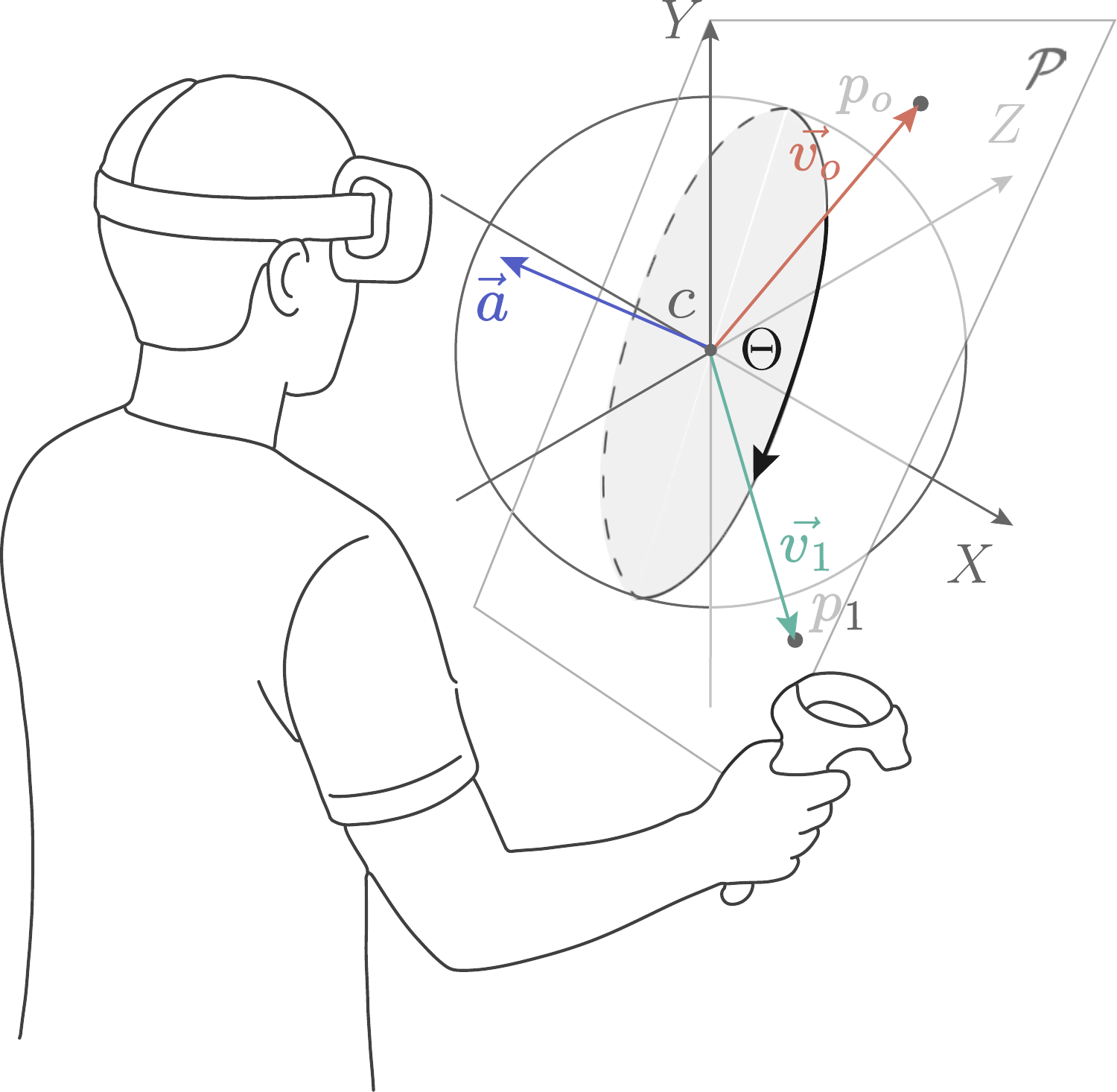}
%\vspace{-0.2cm}
\caption{To rotate the reservoir model, we compute the geodesic arc on $\mathbb{S}^2$ connecting two controller positions---those at the moments of press \( p_0 \) and release \( p_1 \in \mathbb{R}^3 \). These positions are projected onto a virtual sphere centered at the model's origin \( c \), resulting in two normalized direction vectors \( \vec{v}_0 \) and \( \vec{v}_1 \). The rotation axis \( \vec{a} \) is computed as the normalized cross product of these vectors,  \( \vec{a} = \widehat{\vec{v}_0 \times \vec{v}_1} \), which defines the axis orthogonal to the plane \( \mathcal{P} \) spanned by the non-colinear pair \( \vec{v}_0 \) and \( \vec{v}_1 \). The angular displacement \( \theta \) between the two vectors is calculated using the arccosine of their dot product as \( \theta = \arccos(\vec{v}_0 \cdot \vec{v}_1) \). Finally, a rotation quaternion \( q(\vec{a}, \theta) \) is constructed from this axis-angle representation and applied to the model's orientation.}
\label{fig:rotation}
%\vspace{-0.7cm}
\end{figure}

%\vspace{-5px}
\subsection{Spatial Transformations via Physical Metaphors}
%\vspace{-3px}

Regarding pose changes to the model, \textit{translation} is triggered when the user presses and holds the right trackpad. This operation directly maps the controller’s motion to the reservoir model, leveraging a physical analogy: to move the object upward, the user moves the controller upward (\textbf{DG\textsubscript{\textit{Real.}}}).
To minimize unnecessary repetitive movements, we incorporate the concepts of inertia and friction to create a ``sliding object'' metaphor. If the controller's initial motion exceeds a defined velocity threshold, translation is initiated: the reservoir moves at a speed equal to the controller’s initial velocity, and continues to translate until the user releases the right trackpad---thus simulating motion across a frictionless surface.

\begin{figure*}[t!]
\centering
\includegraphics[width=1\linewidth]{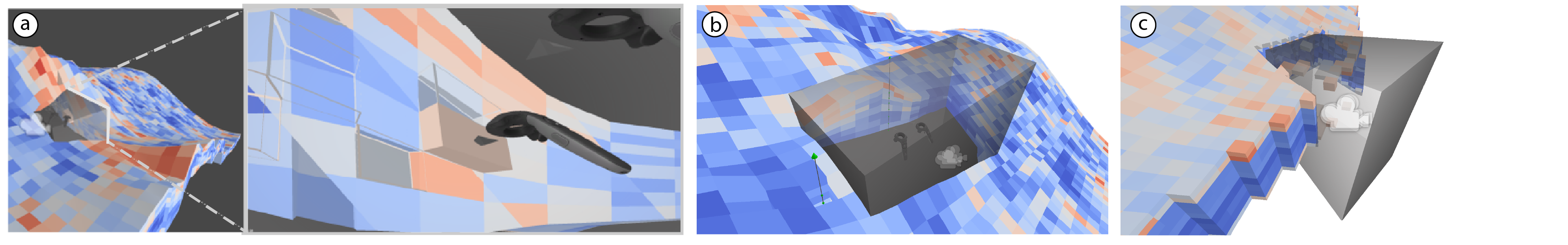}
%\vspace{-0.2cm}
\caption{First-person and bird's-eye views of users performing bimanual single-cell selection using the cutaway lens. Users can adjust contrasting spatial scales to reveal either narrow (a, c) or wide (b) reservoir cross-sections, based on the desired level of structural detail. 
}
\label{fig:cutawayLens}
%\vspace{-0.7cm}
\end{figure*}

\noindent \textit{Rotation} is initiated when users press and hold the left trackpad, as shown in Fig. \ref{fig:rotation}. Rather than mirroring the controller's full 6-DOF orientation, we adapt the 2D Arcball technique to 3D interaction spaces---a method widely used in desktop environments for its strong kinesthetic coherence between cursor movement and object rotation \textcolor{black}{~\cite{Shoemake:1992}}.
Our Arcball-based design not only leverages users’ familiarity with the technique but also trades off some expressiveness in favor of control stability and learnability.
In terms of learnability, our design interprets rotations as the user moving their hand along  a virtual sphere centered at the model, aligning with mental models of rotating physical objects---\textit{e.g.}, turning a volleyball mid-air (\textbf{DG\textsubscript{\textit{Real.}}}).
With respect to stability, direct mirroring can lead to erratic object motion due to unintended hand jitter. In contrast, our approach computes the rotation between two spatial positions (press and release), enabling more deliberate and fine-grained rotations through an arc-based interaction \textbf{DG\textsubscript{\textit{VR.}}}. 
Regarding orientation stability, our method avoids gimbal lock and preserves a continuous orientation, ensuring fluid rotational behavior.
As with translation, the rotation mechanism also incorporates a sliding metaphor to maintain consistency in interaction design and support intuitive physical engagement.

\textit{Scaling} takes advantage of a universal gesture: to spread out the arms represents greater amount; while the opposite movement, less (\textbf{DG\textsubscript{\textit{Real.}}}). Thus, the reservoir model and any other internal entity are scaled up/down whenever users \textit{i}) press and hold both controllers' trackpads simultaneously, and \textit{ii}) open/close their arms (Fig. \ref{fig:gestures}). Although a sliding analogy does not apply to scaling an object, we believe that by maintaining consistency between this metaphor and the two previous ones may easily evoke associations due to physical motion similarities.

%------------------------------------------------
% Hand-anchored Interface Panel
%\vspace{-5px}
\subsection{Hand-anchored Interface Panel}
%\vspace{-3px}

The 1-DOF control panel supports both numerical and categorical variables. Controls include loading geological realizations, and setting filtering thresholds and clustering parameters. The menu is implemented as a closed-loop 3D carousel, where a sequence of rotatable interface items is presented one at a time---allowing multiple menu entries to share a single, compact interaction space.
The body-relative carousel appears near the left controller when its menu button is pressed, and item selection is performed using the right trigger. The carousel operates in the coordinate system of the left controller: the \textit{y}-axis serves as the rotation axis, while the \textit{xy}-plane segments the carousel into two halves---front and back. Pressing the left lateral button rotates the carousel, moving the next item from the back to the front. Items fade out and in as they leave or enter the visible front area.
Furthermore, we provide a kinesthetic 1:1 mapping between the interface panel and the left controller, evoking the experience of wearing a device on the forearm, such as a wristwatch (\textbf{DG\textsubscript{\textit{Real.}}}). This embodied approach addresses two well-known challenges in 3D interface design\textcolor{black}{~\cite{mine1997}}: information retrieval and visual clutter. For the former, leveraging proprioception facilitates quick, intuitive access and even enables eyes-off interaction (\textbf{DG\textsubscript{\textit{VR.}}}). For the latter, placing the panel close to the body minimizes its competition with the primary workspace for visual attention---a critical consideration in VR displays with limited fields of view, such as head-mounted displays (\textbf{DG\textsubscript{\textit{Real.}}}).

\begin{figure}[b!]
\centering
\includegraphics[width=1\linewidth]{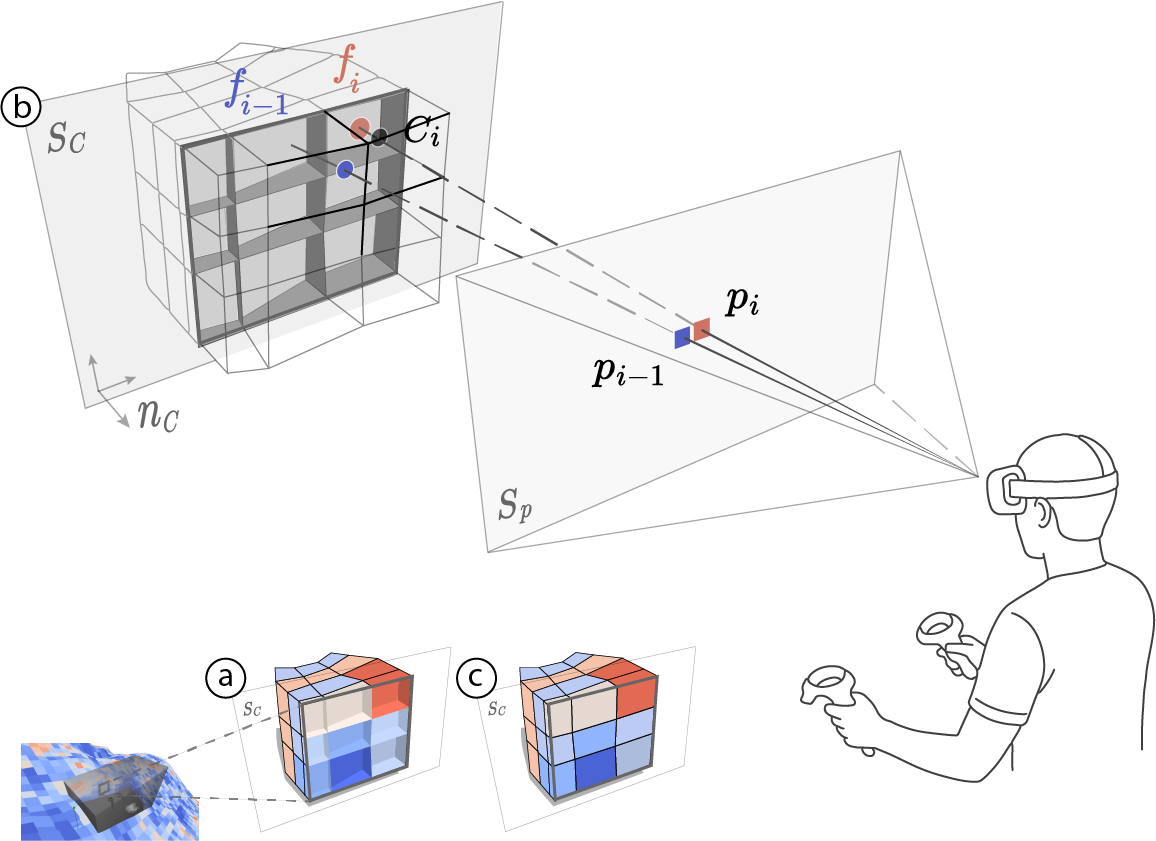}
%\vspace{-0.2cm}
\caption{
Illustrations showcasing the view-dependent cutaway lens, which reveals the reservoir model's internal structures. Non-VOI cells intersected by the cut surface $\mathcal{S}_{c}$ are subject to fragment culling. Discarding fragments that lie in front of $\mathcal{S}_{c}$
results in a hollow-out visual effect of the grid geometry \textcolor{black}{(a)}.
Per-fragment visibility is determined by computing a signed point-plane distance $d(f_{i}, C_{i})$ between a fragment’s projected position $f_{i}$ and the corresponding depth value of the cut surface $C_{i}$ \textcolor{black}{(b)}. Fragments for which $d < 0$ (e.g., $f_{i-1}$) are culled, while those with $d \geq 0$ (e.g., $f_{i}$) are retained. 
To convey the appearance of volumetric solidity in the clipped geometry, surface normals of back-facing fragments are overridden with the local normal of the cut surface, $n_{C}$, at the corresponding fragment. This shading reconstruction yields the perceptual effect of solid elements being cleanly sliced \textcolor{black}{(c)}.
}
\label{fig:cutawayLensSketch}
%\vspace{-0.7cm}
\end{figure}

%------------------------------------------------
% View-dependent Cutaway Lens
%\vspace{-5px}
\subsection{View-dependent Cutaway Lens}
%\vspace{-3px}

%\begin{comment}

Throughout the analysis process, users can enable a \textit{cutaway lens} to address the inherent occlusion caused by the volumetric nature of reservoir grids.
It enables engineers to physically navigate and inspect the internal parts of the reservoir grid---for uncertainty assessment, particularly, they can search for channel structures with great extent of variance.
The lens has the form of a frustum centered on the user’s head, with a kinesthetic isomorphic mapping applied to both translations and rotations---ensuring that head movements is closely mirrored to the visually perceived motion of the frustum.
The lens continuously clips all non-VOI grid cells residing within its interior volume, as illustrated in Fig. \ref{fig:cutawayLens}. By doing so, this culling mechanism produces a projected, non-axis-aligned cross-sectional view that dynamically responds to the user’s head and full-body movements.
This design leverages users’ natural motor abilities to enhance task performance by enabling the combination of multiple actions: continuous head and full-body movements control the sectioning graphics, while the hands remain available to perform complementary tasks---\textit{e.g.}, cell selection (\textbf{DG\textsubscript{\textit{VR.}}}).

Moreover, to maintain interactive frame rates, our cutaway rendering technique is inspired by the method introduced by de Carvalho \textit{et al.} (\textbf{DG\textsubscript{\textit{Scale.}}}) \cite{2016:ICO}.
Similar to their approach, ours leverages screen-space GPU techniques in which volumetric intersections are computed on a per-fragment basis, enabling precise slicing of corner-point grid cells---see Fig. \ref{fig:cutawayLensSketch}. In addition, back-facing surface normals are recomputed  to give the impression of sliced cells appear as solid, rather than hollow, structures.
However, our approach differs in a central aspect: rather than relying on global post-computation of a cut volume, we predefine a local cut geometry that is spatially registered to the user's head-tracked position and view direction. Only the non-VOI cells within this volume are subject to slicing, ensuring that VOI cells remain visible from any viewpoint.
After rendering the cut, the lens geometry is displayed using gradient transparency and rim lighting thus providing depth cues at its boundaries while minimizing visual occlusion.
Our visual design incorporates established principles for reservoir cutaways often employed by geological illustrators: it uses illumination models and simplified cut geometries in order to enhance spatial understanding on intricate corner-point grids \textcolor{black}{\cite{lidal2012design}}.

%------------------------------------------------
% Gesture-based VOI Selection
%\vspace{-5px}
\subsection{Gesture-based VOI Selection}
%\vspace{-3px}

There are two ways of defining the VOI in the variance
model: group and single cell selection. The duration of the trigger press determines the selection type: a brief press-and-release selects a single cell, while a prolonged press-and-hold initiates group selection. 
In the latter case, a hexahedron-shaped selection volume appears at the controller’s current pose. This container continuously grows and mirrors the controller’s motion while the trigger is held---somewhat resembling an inflating balloon (\textbf{DG\textsubscript{\textit{Real.}}}). Upon release, all reservoir cells enclosed within the volume are selected. This design unifies multiple operations---creation, scaling, and placement---into a single, fluid gesture by leveraging users’ proprioception and embodied spatial reasoning (\textbf{DG\textsubscript{\textit{VR.}}}). The selection volume also features cone-shaped handles that can be grabbed and manipulated to modify its geometry.
Additionally, a selection volume can be deleted using a ``throw-away'' metaphor \textbf{DG\textsubscript{\textit{Real.}}}: by grabbing the selection container, performing a swift throwing gesture, and releasing it, the system detects the motion’s velocity. If it surpasses a defined threshold, the container spins away with corresponding linear and angular momentum before disappearing after a brief delay---see Fig. \ref{fig:gestures}. This gesture draws on a universally understood physical metaphor for discarding objects (\textit{e.g.}, crumpling and tossing a sheet of paper), and is designed to be fluid and low-effort---potentially allowing the action to be performed without visual focus on the object being deleted (\textbf{DG\textsubscript{\textit{VR.}}}).

Furthermore, users can select or deselect individual cells to fine-tune previously defined group selections. A brief press-and-release of the trigger button selects the cell currently hovered by the controller’s cursor. Humans' two-handed capabilities allow these operations to be performed using either one or both controllers (\textbf{DG\textsubscript{\textit{VR.}}}).
Together, the selection techniques provide varying levels of granularity, enabling users to construct and iteratively refine VOIs as needed. This versatility is particularly valuable, as highlighted by engineers collaborating with our team, since the criteria for defining regions of interest depend on multiple factors, and precision requirements often vary on a case-by-case basis.

%------------------------------------------------
% Body-relative Clustering Graph
%\vspace{-5px}
\subsection{Body-relative Clustering Graph}
%\vspace{-3px}

The user-defined VOI is one of the primary input parameters for the clustering algorithm, which aims at identifying groups of geo-realizations that potentially yield similar flow simulation behaviors. Its output is a spatial graph, contained within a semi-transparent bounding volume. This ``clustering container'' is body-relative and can be repositioned through direct manipulation, allowing users to place it in the environment relative to their viewpoint
This spatially-anchored control offers advantages in both decluttering and accessibility. Users may reposition the container entirely out of view---such as behind the head---to minimize visual interference with the primary workspace. Conversely, placing it within immediate reach provides convenient access when needed. As body-relative locations are intuitive reference points for spatial memory due to human's sense of proprioception (\textbf{DG\textsubscript{\textit{VR.}}}) \textcolor{black}{\cite{chen2012extending}}, this design enables users to maintain attention on the primary task area even when toggling the clustering container between on- and off-focus states to prevent it from competing for screen space.
Furthermore, prior research provides evidence that individuals are well-adapted to tasks involving asymmetric bimanual coordination, where the non-dominant hand coarsely manipulates a contextual element while the dominant hand performs more precise actions within that context \textcolor{black}{\cite{Kabbash:1993}}. Our system builds on this concept to leverage users’ natural two-handed capabilities: the non-dominant hand positions the graph container at a coarse level, while the dominant hand performs fine-grained operations---hovering and selecting nodes (\textbf{DG\textsubscript{\textit{VR.}}}).

%-------------------------------------------
% User Study
%-------------------------------------------
%\vspace{-6px}
\section{\textcolor{black}{User Study}}
\label{sec:userStudy}
%\vspace{-3px}

This section presents user study conducted with 12 reservoir engineers from an industry partner. Given the domain-specific character of our application, the limited scope of the study was appropriate. The primary goals of the sessions were to assess engineers’ perceptions of our design choices, gather suggestions for refinement and future research. 
It is important to note that, since our tool enables a novel, user-driven approach to uncertainty analysis, a strict comparative evaluation against commercial software was not feasible. Instead, we designed a task-oriented study based on a realistic case scenario, following the workflow outlined in Section \textcolor{blue}{~\ref{sec:systemDataModel}}. Details of the evaluation are presented below.

\noindent \textbf{Participants} The study included 12 participants: 8 male and 4 female, aged between 24 and 58 years. All participants held one or more graduate degrees in petroleum engineering and had between 1 and 7 years of professional experience in reservoir analysis.

\noindent \textbf{Method} Each evaluation session lasted about 90 minutes, and was audio and video recorded for subsequent qualitative analysis. Sessions began with a brief overview of the study’s objectives. Participants then completed a pre-questionnaire to gather personal information---\textit{e.g.} academic and professional credentials, age, and comfort level with emerging technologies. This was followed by an introduction to the uncertainty analysis process supported by our system.
Following this initial orientation, participants completed a series of four task assignments. Each consisted of  \textit{i}) a demonstration tutorial by the researcher, \textit{ii}) a short trial period for free exploration, \textit{iii}) a set of goal-directed analysis tasks for participants to perform, and \textit{iv}) a semi-structured interview to reflect on their experience.
Finally, upon completing the task assignments, participants were asked to complete a post-questionnaire designed to assess perceived ease of use, ease of learning, usefulness, and satisfaction with the system \textcolor{black}{~\cite{Davis:1989}}. The items from the USE questionnaire were randomized to reduce potential response bias. Since the adoption of innovative technology is strongly influenced by users’ perceptions of their usefulness and ease of use, these constructs offer valuable evidence as to whether our design aligns with user anticipations and supports the intended workflow.

\noindent \textbf{Tasks} An inquiry-based learning approach\textcolor{black}{~\cite{alesandrini:2002}} was used throughout the tasks, with the researcher posing guiding questions and encouraging participants to explore and reflect on their findings. A total of 21 tasks were included, each allowing for multiple solution strategies---an approach aimed at reinforcing concepts, promoting engagement, and evaluating the system’s learning curve.
The first task assignment covered interpretation of geological variability for identifying reservoir regions with likely dissimilar flow behavior, with the researcher providing questions such as: “\textit{can you move closer to a well with a reasonable high-variance permeability surrounding?}” or “\textit{can you identify a low-variability flow channel?}”
The second assignment involved locating and specifying a VOI. Participants were asked questions such as: “\textit{can you specify cell groups with low variability within this injection-production well pattern?}”
The third and fourth assignments addressed clustering-based analysis. For the third, participants were provided with a predefined VOI and corresponding cluster graph. The tasks focused on identifying outliers as participants were prompted with questions such as: “\textit{can you tell me the distance between model 9 and its cluster center?}” and “\textit{can you identify the outermost outlier in this cluster?}”
The final assignment built upon the previous one, asking participants to make subjective decisions regarding outlier inclusion in the default representative set. The researcher prompted questions such as “\textit{would you consider adding new representative models? why?}”

%-------------------------------------------
% Results \& Discussion
%-------------------------------------------
%\vspace{-6px}
\section{\textcolor{black}{Results \& Discussion}}
\label{sec:resultsDiscussion}
%\vspace{-3px}
This section deliberates users' responses towards our system regarding our design choices, discussing perceived strengths, limitations, and opportunities for future refinement of our tool.
In addition, we catalog our findings as a set of design recommendations intended to guide researchers and tool-builders of immersive interfaces---both within reservoir engineering and broader application domains.

\begin{figure}[t!]
%\centering
\includegraphics[width=0.98\linewidth]{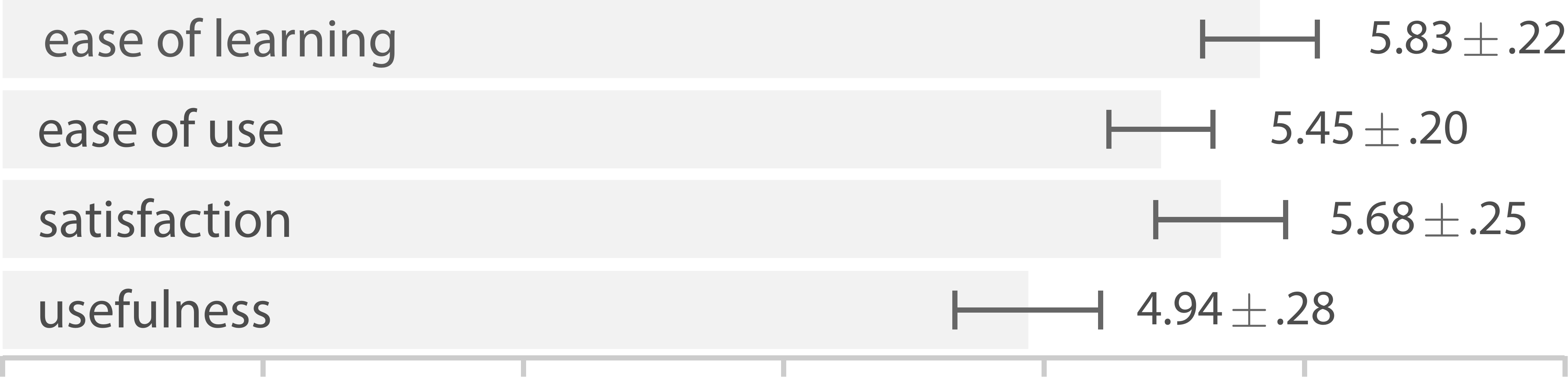}
%\vspace{-0.2cm}
\caption{Average scores from the USE questionnaire. Error bars indicate standard error.}
\label{fig:studyResults}
%\vspace*{-6px}
\end{figure}

\noindent \textbf{Quantitative} The results of our study are encouraging. Participants were able to independently navigate the system and became comfortable and proficient in completing most assignments. USE questionnaire results confirmed that users found the system easy to learn and the interfaces pleasant to use, while feeling capable of accomplishing their intended goals. \textcolor{black}{Figure~\ref{fig:studyResults}} presents the average scores across all questionnaire dimensions, which demonstrated high internal consistency (\textit{Cronbach’s alpha $>$ 0.80}).
Notably, the ``usefulness'' dimension received a lower average score and exhibited greater disagreement among participants ($4.94 \pm 0.28$). Audio recordings from interviews provided some insight into this finding. When asked about the system’s practical suitability, participants acknowledged its potential but emphasized portability as a key requirement. They pointed out that workflows in the oil and gas industry involve multiple interdependent tools, necessitating the ability to save and transfer data between systems. We recognize this limitation and consider software portability an important direction for future development.

\noindent \textbf{Qualitative} Qualitative data also revealed valuable insights into interface elements that could inform future VR designs, both within petroleum engineering and other domains. To abstract this data into design opportunities for future research, we transcribed each user session---including video, audio, and notes---and produced written summaries listing key discussion points. These were broadly grouped according to the four task assignments. Within each group, summary points were analyzed and categorized into high-level themes. As relevant topics emerged, the corresponding recordings were revisited to extract meaningful behaviors and illustrative quotes. In what follows, the outcomes of this analysis are discussed as a set of design recommendations.

\noindent \textit{Support bimanual interaction to distribute task load}.
Observations and video recordings showed that the majority of participants (11 out of 12) predominantly used two-handed interactions during tasks---\textit{e.g.}, holding the cluster container with one hand while inspecting outliers with the other. This frequent use of compound interactions aligns well with our design goals and suggests an effective distribution of task functions across both hands, without imposing significant cognitive or motor load.

\noindent \textit{Reinforce visual cues for reference frame and operation mechanics}.
Regarding our gesture-based designs, translation and scaling were judged very intuitive by all participants. However, four reported initial confusion with rotation. Observation indicated this stemmed from differing assumptions about the frame of reference---some expected an egocentric frame (common in VR), while our system used an allocentric one (typical in CAD software). Therefore, a key design takeaway is to reinforce visual cues to clarify the reference frame and operation mechanics, solidifying users’ mental models---particularly during the initial learning phase.

\noindent \textit{Motor-based interfaces should support diverse skill levels.}
The nonlinear mappings and sliding metaphors had human factors implications. By relying on motor control, these interfaces enabled participants to adopt strategies aligned with their degree of comfort and proficiency---\textit{e.g.}, when managing speed-accuracy trade-offs, some novice users deliberately used small, repetitive gestures for precise positioning. Such design strategies seems to support skill development by accommodating a range of expertise levels and promoting the transition from novice to expert.

\noindent \textit{View-responsive lenses should account for unintentional head shifts.}
Participants were most enthusiastic about the cutaway capability, which inspired redesign suggestions. Six participants mentioned the sensitivity of the lens to head movement. As one explained: ``\textit{there were times when I was working on a particular cross-section, picking cells, and a slight head movement shifted the view (...) it took some time to find that specific section again.}'' Similar behavior was observed in video recordings of other participants, even if not explicitly mentioned in their interviews. To address this issue, users recommended implementing a feature to “lock” or “anchor” the focused views during task execution.

\noindent \textit{Support conformal or composite lensing for more irregular geometry}.
Three participants highlighted the usefulness of cutaway lenses for examining well trajectories. As one noted, ``\textit{for vertical wells, using cross-sections may be good enough}''; however, for non-conventional wells, ``\textit{it is challenging to make sense of data across multiple sections}.'' In response to this challenge, a participant suggested a mechanism to enable users combine multiple cross-sections during the design of a new well path, allowing them to later revisit reservoir behavior along those sections to assess potential oil production. The same participant also remarked that this functionality would be valuable ``\textit{for defining fence diagrams to examine stratigraphic and structural formations between wells}''.
Additional suggestions included the ability to sculpt deformable sections or adjust the lens shape based on user needs. These features could significantly enhance front-tracking visualizations in fluid simulations. As one participant explained: ``\textit{when performing water flooding, for example, we need to track the water front without losing reference to the rest of the reservoir model (...) to find regions with higher mobility (...) but because the front is dynamic, it can change shape (...) depending on saturation, pressure, etc.}''

\noindent \textit{Align button controls with natural physical analogies whenever possible}.
Regarding VOI selection, five participants reported that the gesture-based volume selection was very easy to learn and use. However, they also noted that it took some time to become familiar with the button mappings. We observed that participants generally performed the correct gestures---\textit{e.g.}, placing the controller inside the container to grab it or enacting a ``throw away'' motion, but often confused the trigger (used for creation) with the trackpad (used for grabbing).
This was an intriguing takeaway: the learning curve for gestures appeared to be shorter than for button mappings. We believe that such confusion could be mitigated, at least in part, by refining the button layout to better align with physical analogies---\textit{e.g.}, using the grip buttons on both sides of the controller to more naturally mimic the action of closing a hand to grab an object. Nonetheless, we acknowledge that hardware limitations may constrain the ability to implement such physically intuitive mappings in varied scenarios.

\noindent \textit{Balance effort with performance in real-world analogies.}
Additionally, although one participant acknowledged the ``throw away'' gesture as intuitive, they questioned the degree of physical effort compared to simply pressing a button as in traditional interfaces. Interestingly, this same participant had previously expressed very positive feedback regarding single selection combined with the cutaway lens---even though that interaction involves continuous head and arm movement, making it more physically demanding.
We hypothesize that this behavior reflects a user-evaluated trade-off between physical effort and perceived gains in task performance. This highlights an important design consideration when incorporating real-world analogies: balancing learning speed, physical load, and performance benefits. While this interpretation is grounded in qualitative analysis, it warrants further investigation through controlled experimentation.

%-------------------------------------------
% Conclusion & Future Work
%-------------------------------------------
%\vspace{-6px}
\section{\textcolor{black}{Conclusion \& Future Work}}
\label{sec:conclusionFutureWork}
%\vspace{-3px}

We introduced a novel VR system for uncertainty analysis of geological models, developed through our close collaboration with reservoir engineers. We described the design goals and the rationale behind our system, whose spatial interfaces incorporates reality-based and fluid interaction paradigms aimed at improving usability.
We also reported a task-driven user evaluation involving open-ended feedback sessions with 12 reservoir engineers from an industry partner. Feedback was collected through notes, audio and video recordings, questionnaires, and semi-structured interviews.
Findings from our study are promising: they support our design goals and understanding of the needs and prospects surrounding the system's role in facilitating the intended interactive framework for geological uncertainty analysis.
Suggestions provided by participants will guide our future work---notably, given their enthusiastic response, we plan to further investigate lens composition to enable more versatile cross-sectional profiles compared to rigid-shaped lenses.
Furthermore, we distilled the study findings into a set of design recommendations which we hope will provide valuable insights for future researchers and developers in designing spatial interfaces for petroleum engineering and other application domains. More broadly, we believe this work demonstrates the validity and potential of 3D user interfaces for scientific data analysis.

%-------------------------------------------------------------------------
% bibtex
\bibliographystyle{eg-alpha-doi} 
\bibliography{egbibsample}       

\newcommand{\etalchar}[1]{$^{#1}$}
\begin{thebibliography}{\uppercase{MFVBCS15}}

\bibitem[AL02]{alesandrini:2002}
\textsc{Alesandrini K., Larson L.}:
\newblock Teachers bridge to constructivism.
\newblock \emph{The clearing house 75}, 3 (2002), 118--121.

\bibitem[AVBSCS14]{Amorim:2014}
\textsc{Amorim R., Vital~Brazil E., Samavati F., Costa~Sousa M.}:
\newblock {3D G}eological modeling using sketches and annotations from geologic maps.
\newblock In \emph{Proceedings of the Symposium on Computational Aesthetics, Non-Photorealistic Animation and Rendering, and Sketch-Based Interfaces and Modeling} (2014), pp.~17--25.

\bibitem[Bar97]{Bartram:1997}
\textsc{Bartram L.}:
\newblock Can motion increase user interface bandwidth in complex systems?
\newblock In \emph{International Conference on Systems, Man, and Cybernetics. Computational Cybernetics and Simulation} (1997), vol.~2, pp.~1686--1692.

\bibitem[BJA92]{Ballin:1992}
\textsc{Ballin P., Journel A., Aziz K.}:
\newblock Prediction of uncertainty in reservoir performance forecast.
\newblock \emph{Journal of Canadian Petroleum Technology 31}, 4 (1992), 52--62.

\bibitem[CMG13]{CMG}
{CMG Software Overview - Computer Modelling Group LTD}.
\newblock \url{http://www.cmgl.ca/software}, Acessed: 2018-04-13.

\bibitem[CMT{\etalchar{*}}12]{chen2012extending}
\textsc{Chen X., Marquardt N., Tang A., Boring S., Greenberg S.}:
\newblock Extending a mobile device's interaction space through body-centric interaction.
\newblock In \emph{Proceedings of the International Conference on Human-Computer Interaction with Mobile Devices and Services} (2012), pp.~151--160.

\bibitem[Csi08]{Csikszentmihalyi:2008}
\textsc{Csikszentmihalyi M.}:
\newblock \emph{Flow: The Psychology of Optimal Experience}.
\newblock Harper Collins, 2008.

\bibitem[Dav89]{Davis:1989}
\textsc{Davis F.~D.}:
\newblock Perceived usefulness, perceived ease of use, and user acceptance of information technology.
\newblock \emph{MIS Quarterly 13}, 3 (1989), 319--340.

\bibitem[dCVBM{\etalchar{*}}16]{2016:ICO}
\textsc{de~Carvalho F.~M., Vital~Brazil E., Marroquim R.~G., Costa~Sousa M., Oliveira A.}:
\newblock Interactive cutaways of oil reservoirs.
\newblock \emph{Graphical Models 84}, C (2016), 1--14.

\bibitem[DGK04]{Dhillon:2004}
\textsc{Dhillon I.~S., Guan Y., Kulis B.}:
\newblock Kernel {K}-means: Spectral clustering and normalized cuts.
\newblock In \emph{Proceedings of the ACM International Conference on Knowledge Discovery and Data Mining} (2004), pp.~551--556.

\bibitem[EVMJ{\etalchar{*}}11]{Elmqvist:2011}
\textsc{Elmqvist N., Vande~Moere A., Jetter H.-C., Cernea D., Reiterer H., Jankun-Kelly T.~J.}:
\newblock Fluid interaction for information visualization.
\newblock \emph{Information Visualization 10}, 4 (2011), 327--340.

\bibitem[FC11]{France:2011}
\textsc{France S.~L., Carroll J.~D.}:
\newblock Two-way multidimensional scaling: A review.
\newblock \emph{IEEE Transactions on Systems, Man, and Cybernetics 41}, 5 (2011), 644--661.

\bibitem[GBK{\etalchar{*}}01]{Grossman:2001}
\textsc{Grossman T., Balakrishnan R., Kurtenbach G., Fitzmaurice G., Khan A., Buxton B.}:
\newblock Interaction techniques for {3D} modeling on large displays.
\newblock In \emph{Proceedings of the Symposium on Interactive 3D Graphics} (2001), pp.~17--23.

\bibitem[Gos12]{Goshtasby:2012}
\textsc{Goshtasby A.~A.}:
\newblock \emph{Image Registration: Principles, Tools and Methods}.
\newblock Springer Science \& Business Media, 2012.

\bibitem[Gru04]{Gruchalla:2004}
\textsc{Gruchalla K.}:
\newblock Immersive well-path editing: investigating the added value of immersion.
\newblock In \emph{IEEE Virtual Reality} (2004), pp.~157--164.

\bibitem[HHN85]{Hutchins:1985}
\textsc{Hutchins E.~L., Hollan J.~D., Norman D.~A.}:
\newblock Direct manipulation interfaces.
\newblock \emph{Human-Computer Interaction 1}, 4 (1985), 311--338.

\bibitem[JGH{\etalchar{*}}08]{Jacob:2008}
\textsc{Jacob R.~J., Girouard A., Hirshfield L.~M., Horn M.~S., Shaer O., Solovey E.~T., Zigelbaum J.}:
\newblock Reality-based interaction: A framework for {post-WIMP} interfaces.
\newblock In \emph{Proceedings of the SIGCHI Conference on Human Factors in Computing Systems} (2008), pp.~201--210.

\bibitem[KB15]{kinsland2015}
\textsc{Kinsland G.~L., Borst C.~W.}:
\newblock Visualization and interpretation of geologic data in {3D} virtual reality.
\newblock \emph{Interpretation 3}, 3 (2015), SX13--SX20.

\bibitem[KMB93]{Kabbash:1993}
\textsc{Kabbash P., MacKenzie I.~S., Buxton W.}:
\newblock Human performance using computer input devices in the preferred and non-preferred hands.
\newblock In \emph{Proceedings of the Conference on Human Factors in Computing Systems} (1993), pp.~474--481.

\bibitem[LHV12]{lidal2012design}
\textsc{Lidal E.~M., Hauser H., Viola I.}:
\newblock Design principles for cutaway visualization of geological models.
\newblock In \emph{Proceedings of the Conference on Computer Graphics} (2012), pp.~47--54.

\bibitem[LLG{\etalchar{*}}07]{Lidal:2007}
\textsc{Lidal E.~M., Langeland T., Giertsen C., Grimsgaard J., Helland R.}:
\newblock A decade of increased oil recovery in virtual reality.
\newblock \emph{IEEE Computer Graphics and Applications 27}, 6 (2007), 94--97.

\bibitem[MCS{\etalchar{*}}16]{Mota:2016}
\textsc{Mota R. C.~R., Cartwright S., Sharlin E., Hamdi H., Costa~Sousa M., Chen Z.}:
\newblock Exploring immersive interfaces for well placement optimization in reservoir models.
\newblock In \emph{Proceedings of the Symposium on Spatial User Interaction} (2016), pp.~121--130.

\bibitem[MFVBCS15]{martins2015}
\textsc{Martins~Filho Z., Vital~Brazil E., Costa~Sousa M.}:
\newblock Exploded view diagrams of {3D} grids.
\newblock In \emph{Conference on Graphics, Patterns and Images} (2015), pp.~242--249.

\bibitem[Min97]{mine1997}
\textsc{Mine M.~R.}:
\newblock {ISAAC: a meta-CAD system for virtual environments}.
\newblock \emph{Computer-Aided Design 29}, 8 (1997), 547--553.

\bibitem[MLP{\etalchar{*}}11]{ma2011}
\textsc{Ma Y.~Z., La~Pointe P.~R., et~al.}:
\newblock \emph{Uncertainty Analysis and Reservoir Modeling: Developing and Managing Assets in an Uncertain World}, vol.~96.
\newblock 2011.

\bibitem[Pet13]{Petrel}
{Petrel E\&P software platform - Schlumberger software}.
\newblock \url{http://www.software.slb.com/products/petrel}, Acessed: 2018-04-13.

\bibitem[RKSB12]{ragan2012studying}
\textsc{Ragan E.~D., Kopper R., Schuchardt P., Bowman D.~A.}:
\newblock Studying the effects of stereo, head tracking, and field of regard on a small-scale spatial judgment task.
\newblock \emph{IEEE Transactions on Visualization and Computer Graphics 19}, 5 (2012), 886--896.

\bibitem[SC10]{Scheidt:2010}
\textsc{Scheidt C., Caers J.}:
\newblock Bootstrap confidence intervals for reservoir model selection techniques.
\newblock \emph{Computational Geosciences 14}, 2 (2010), 369--382.

\bibitem[SCSCS14]{Somanath:2014}
\textsc{Somanath S., Carpendale S., Sharlin E., Costa~Sousa M.}:
\newblock Information visualization techniques for exploring oil well trajectories in reservoir models.
\newblock In \emph{Proceedings of Graphics Interface} (2014), pp.~145--150.

\bibitem[SHM{\etalchar{*}}16]{sahaf2016}
\textsc{Sahaf Z., Hamdi H., Maurer F., Nghiem L., Costa~Sousa M.}:
\newblock Clustering of geological models for reservoir simulation studies in a visual analytics framework.
\newblock In \emph{European Association of Geoscientists and Engineers Conference and Exhibition} (2016).

\bibitem[SHM{\etalchar{*}}18]{Sahaf:2018}
\textsc{Sahaf Z., Hamdi H., Mota R. C.~R., Costa~Sousa M., Maurer F.}:
\newblock A visual analytics framework for exploring uncertainties in reservoir models.
\newblock In \emph{Proceedings of the Conference on Information Visualization Theory and Applications} (2018), pp.~74--84.

\bibitem[Sho92]{Shoemake:1992}
\textsc{Shoemake K.}:
\newblock {ARCBALL: A user interface for specifying three-dimensional orientation using a mouse}.
\newblock In \emph{Proceedings of the Conference on Graphics Interface} (1992), pp.~151--156.

\bibitem[Som12]{Somanath:Thesis:2012}
\textsc{Somanath S.}:
\newblock \emph{{Exploring Tabletops as an Interaction Medium in the Context of Reservoir Engineering}}.
\newblock Master's thesis, University of Calgary, Canada, 2012.

\bibitem[SSSCS11]{Sultanum:2011}
\textsc{Sultanum N., Somanath S., Sharlin E., Costa~Sousa M.}:
\newblock "{P}oint it, split it, peel it, view it": Techniques for interactive reservoir visualization on tabletops.
\newblock In \emph{Proceedings of the ACM International Conference on Interactive Tabletops and Surfaces} (2011), pp.~192--201.

\end{thebibliography}

% biblatex with biber
% \printbibliography                

%-------------------------------------------------------------------------
%Color tables are no longer required for purely electronic publications.
\newpage

\end{document}